\documentclass[10pt,letterpaper,twocolumn]{article}
\usepackage[latin1]{inputenc}
\usepackage{amsmath}
\usepackage{amsfonts}
\usepackage{amssymb}
\usepackage[dvips]{graphicx}
\author{A. A. Arellano-Baeza(a), R. V. Garcia(b), and M. Trejo-Soto(b)
\\
\\ (a) Mining Department, University of Santiago de Chile 
\\ (b) Earth Sciences School, Autonomous University of Sinaloa}
\title{Use of high resolution satellite images for tracking of changes in the lineament structure, caused by earthquakes.}
\begin{document}
\maketitle

\begin{abstract}
Over the last decades strong efforts have been made to apply new spaceborne technologies to the study and possible forecast of strong earthquakes. In this study we use ASTER/TERRA multispectral satellite images for detection and analysis of changes in the system of lineaments previous to a strong earthquake. A lineament is a straight or a somewhat curved feature in an image, which it is possible to detect by a special processing of images based on directional filtering and or Hough transform. "The Lineament Extraction and Stripes Statistic Analysis" (LESSA) software package, developed by Zlatopolsky (1992,  1997). We assume that the lineaments allow to detect, at least partially, the presence ruptures in the Earths crust, and therefore enable one to follow the changes in the system of faults and fractures associated with strong earthquakes. We analysed 6 earthquakes occurred in the Pacific coast of the South America and one earthquake in Tibet, Xizang, China with the Richter scale magnitude $\geq 5.2$ Mw. They were located in the regions with small seasonal variations and limited vegetation to facilitate the tracking of features associated with the seismic activity only. It was found that the number and orientation of lineaments changed significantly about one month before an earthquake approximately, after that the system  gradually returns its initial state. This effect increases with the earthquake magnitude, and it is much more easily detectable in case of convergent plate boundaries (for example, Nazca and South American plates). The results obtained open a possibility to develop a methodology able to evaluate the seismic risk in the regions with similar geological conditions.
\end{abstract}

\section{Introduction}
Throughout the world, devastating earthquakes constantly occur with little or no advance warning. Brune (1979) proposed that earthquakes may be inherently unpredictable since large earthquakes start as smaller earthquakes, which in turn start as smaller earthquakes, and so on. In his model, most of the fault is in a state of stress below that required to initiate slip, but it can be triggered and caused to slip by nearby earthquakes or propagating ruptures. Any precursory phenomena will only occur when stresses are close to the yield stress. However, since even small earthquakes are initiated by still smaller earthquakes, in the limit, the region of rupture initiation where precursory phenomena might be expected is vanishingly small. Even if every small earthquake could be predicted, one still faces the impossible task of deciding which of the thousands of small events will lead to a runaway cascade of rupture composing a large event.

    Nevertheless, the discussion about the possibility of earthquake forecast continues to be open, and a wide spectrum of new spaceborne technologies for the earthquake study and forecast appeared during the last decades. The main advantage of spaceborne technologies is the ability to cover big territories  and areas with difficult access. The list of these technologies is very large. As an example it is possible to mention the measurements of different ionospheric precursors of earthquakes including changes in electromagnetic ELF radiation (Serebryakova et al., 1992, Gokhberg et al., 1995), and ionospheric electron temperature (Sharma et al., 2006), and density (Trigunait et al., 2004) (see Pulinets at al., (2003) for a review).  Many efforts have been concentrated in the study of the ground deformation using the satellite radar interferometry, that makes it possible to determine the location and amount of coseismic surface displacements (see for example Satybala, 2006; Schmidt and Bergmann, 2006, Lasserre et al., 2005, Funning et al., 2005). The IR satellite thermal imaging data were used to study pre-earthquake thermal anomalies (Ouzounov and Freund, 2004). The anomalies in the surface latent heat flux data were also detected a few days prior to coastal earthquakes (Cervone et al., 2005, Singh and Ouzounov, 2003; Dey at al., 2004).

    During last years, significant progress has been reached in the understanding how the complex set of phenomena, related to the earthquake gestation is reflected, at least partially, in the geological lineaments. In particular, Cotilla-Rodriguez and Cordoba-Barba (2004) studied the morphotectonic structure of the Iberian Peninsula and showed that the main seismic activity is concentrated on the first- and second rank lineaments, and some of important epicenters are located near the lineament intersections. Stich et al., (2001) found from the analysis of 721 earthquakes with magnitude between 1.5 and 5.0, that the epicenters draw well-defined lineaments and show two dominant strike directions N120-130E and N60-70E, which are coincident with the known fault system of the area. Distances within multiplets (typically several tens of meters) are smaller than the fracture radii of these events.  Carver et al. (2003) have used the SRTM and Landsat-7 digital data and paleoseismic techniques to identify active faults and evaluate seismic hazards on the northeast coast of Kodiak Island, Alaska. 

    Arellano et al. (2004ab, 2005) studied the changes in the lineament structure caused by a 5.2 Richter scale magnitude earthquake occurred January 27, 2004 in southern Peru. During last years this region is studied intensively using the ground based seismic network (Comte et al., 2003; David et al., 2004; Legrand, 2005) as well as GPS and SAR interferometry data (Campos et al., 2005). The ASTER/TERRA high resolution multispectral images 128 and 48 days before and 73 days after the earthquake were used. It was shown that the lineament system is very dynamical, and significant numbers of lineaments appeared between four and one month before the earthquake. They also studied the changes in stripe density fields. These fields represent the density of stripes, calculated for each direction as a convolution between the corresponding circular masks and the image. The stripe density field residuals showed the reorientation of stripes, which agrees with the dilatancy models of earthquakes. These features disappear in the image obtained two months after the earthquake. Analysis of the similar reference area, situated at 200 km from the epicenter, showed that in the absence of earthquakes both lineaments and stripe density fields remain unchanged. Similar results were obtained later by Bondur and Zverev (2005) due to analysis of MODIS (TERRA) images of earthquake in California.

    Singh V.P and R.P. Singh (2005) used the lineament analysis to study changes in stress pattern around the epicenter of Mw=7.6 Bhuj earthquake. This earthquake occurred 26 January 2001 in India. Indian Remote Sensing (IRS-1D) LISS data were used. The lineaments were extracted using high pass filter (Sobel filter in all directions).  The results obtained also confirm that the lineaments retrieved from the images 22 days before the earthquake differ from the lineaments obtained 3 days after the earthquake. It was assumed that they are related to fractures and faults and their orientation and density give an idea about the fracture pattern of rocks. The results also show the high level of correlation between the continued horizontal maximum compressive stress deduced from the lineament and the earthquake focal mechanism.

    Studies of lineament dynamics can also contribute to better understanding of the nature of earthquakes. To date significant number of theories has been developed to explain how an earthquake occur. One of the oldest is the elastic rebound theory, proposed by Harry Reid after the California 1906 earthquake (Reid, 1910). It is based on the assumption that the rocks under stress deform elastically, analogous to a rubber band. Strain builds up until either the rock break creating a new fault or movement occurs on an existing fault. As stored strain is released during an earthquake, the deformed rocks "rebound" to their undeformed shapes. The magnitude of the earthquake reflects how much strain was released. The seismic gap hypothesis states that strong earthquakes are unlikely in regions where weak earthquakes are common and the longer the quiescent period between earthquakes, the stronger the earthquake will be when it finally occur (see Kagan and Jackson, 1995, and references therein). The complication is that the boundaries between crustal plates are often fractured into a vast network of minor faults that intersect the major fault lines. When an earthquake relieves the stress in any of these faults, it may pile additional stress on another fault in the network. This contradicts the seismic gap theory because a series of small earthquakes in an area can then increase the probability that a large earthquake will follow. 

    The theory of dilatancy states that an earthquake develops similarly to the rupture of a solid body (Whitcomb et al., 1973; Scholz et al., 1973; Griggs et al., 1975). This approach has a physical basis in laboratory studies of rock samples, which showed that when rocks are compressed until they fracture, a dilatancy often occurs for a short time interval immediately before failure (Scholz, 1968). Mjachkin et al. (1975ab) modified the dilatancy approach and formulated the theory of unstable avalanche crack formation. The model is based on the two phenomena: interaction between the stress fields of the cracks, and the localization of the process of the crack formation. The number and size of cracks increases gradually under the action of tensions below a critical value. When the density of cracks reaches some critical value, the rock breaks very quickly. This process develops due to merging of cracks as a result of interaction between their stress fields. However, the larger cracks have more probability to interact, and it supposes that a small number of large cracks is gradually formed, and their merging leads to the macro-destruction. During the earthquake gestation, a gradual increase in number and size of cracks occur in the whole volume of rock  under compression. When the crack density reaches a critical value, the barriers between cracks are destroyed, and the velocity of deformation increases. Finally, an unstable deformation develops and localizes in a narrow zone of future macro-rupture, the cracks orient along the future macro rupture, and  a macro-crack is formed, producing an earthquake. 

    However, this model was modified recently by introducing a concept of self-organized criticality, proposed by Bak et al. (1988) for description of the behavior of complex systems.  Applied to earthquakes, this approach describes an interaction between the ruptures of different rank and the collective effects of rupture formation before a strong earthquake (for example Varnes, 1989; Keilis-Borok, 1990; Sammis and Sornette, 2002). A wide area around the future epicenter reaches a metastable state, and the system turns to be very sensitive to small external actions. The concept of SOC does not contradict the concept of dilatancy. However, it assumes that a significantly greater region is involved during the last stages of earthquake preparation than the dilatancy theories imply. 

    Unfortunately, the main processes leading to an earthquake develop deep inside the crust, and there is no way to realize direct measurements of any quantity. The unique possibility we have is to search for traces of these processes disseminated over the Earth's surface. In this context, the lineament analysis could convert in the future in one of power tools for earthquake study, complementing other ground-based and satellite studies. Nevertheless, despite promising results obtained, many important questions continue to be present. It is necessary to understand, whether the lineament system is always affected by earthquake? How early before an earthquake is this alteration manifested? How is it related to the earthquake magnitude and depth? How different is it in case of different kinds of plate borders? This study represent a first step in the search of some answers.
    
\section{Instrumentation and Data Analysis}

For this study we used the the images from the Advanced spaceborne Thermal Emission and Reflection Radiometer (ASTER) onboard the TERRA satellite. The satellite was launched to a circular solar-synchronous orbit with altitude of 705 km. The radiometer is composed by three instruments: Visible and Near Infrared Radiometer (VNIR) with 15 m resolution (bands 1-3), Short Wave Infrared Radiometer (SWIR) with 30 m resolution (bands 4-9) and Thermal Infrared Radiometer TIR with 90 m resolution (bands 11-14) which measure the reflected and emitted radiation of the Earths surface covering the range 0.56 to 11.3 $\mu$m (Abrams, 2000). 

    The images were processed using the Lineament Extraction and Stripes Statistic Analysis (LESSA) software package (Zlatopolsky, 1992, 1997), which provides a statistical description of the position and orientation of short linear structures through detection of small linear features (stripes) and calculation of descriptors that characterize the spatial distribution of stripes. The program also makes it possible to extract the lineaments  - straight lines crossing a significant part of the image. To make this extraction, a set of very long and very narrow (a few pixels) windows (bands), crossing the entire image in different directions, was used. In each band the density of stripes, the direction of which is coincident with the direction of the band, is calculated. When the density of stripes overcomes a pre-established threshold, the chain of stripes along the band is considered as a lineament. The value of threshold depends on the brightness of the image, relief, etc. and is established empirically. Previous studies showed that lineaments, extracted from the image by applying the LESSA program, are strongly related to the main lineaments, obtained from the geomorphological studies (Zlatopolsky, 1992, 1997). The details about the application of LESSA package for earthquake studies is given in (Arellano et al., 2005).

    During this study we analysed 5 earthquakes, occurred in the in the Pacific coast of the South America and one earthquake occured in Himalaya, China. Table 1 resumes main characteristics of these earthquakes, indicating the date, country, geographic coordinates, magnitude, and depth of the earthquake. Also the the ASTER images available for each earthquake are  indicated, for example -126 means that the image 126 days before earthquake was used. The last column indicates that in all South American earthquakes number and orientation of lineaments suffered changes before the earthquake. In case of China earthquake, we can not give a defnite answer, because unfortunately the key images tens day before the earthquake were covered by clouds in appoximately 50\%, that made the lineament analysis difficult (last two lines, two areas cvering the hipocenter and close to hipocenter). Neverthless, more sophysticated technique  based on analysis of stipe density fields was able to detect the alterations in these fields related to the earthquake. The methodology of this analysis is given in (Arellano et al., 2006). Currently we are preparring a manuscript dedicated especially to the analysis of this event. 

To illustrat the results obtained we give as an example a detailed analysis of 7.8 Mw earthquake, which took place June 13, 2005 in  northern Chile close to Arica (see Figure 1). The hipocenter was situated at 115 km deep in the crust. The coordinates were $-19.99{\circ}$ LAT, $-69.197{\circ}$ LONG. In the top, a series of four band 3 ASTER (VNIR) images around the hipocenter area are shown. It is possible to see, that the presence of clouds was low. The second line contains the images showing the systems of lineaments, obtainend from the images above using the LESSA programm with a threhhold 120 (Zlatopolsky, 1992, 1997). It is posible to see clear time evolution of lineaments, experimenting strong increase in the number of lineaments 5 days before the earthquake. The third and fourth lines quanify this effect by calculaing the rose-diagramms and the Radon transforms. Reorientation of lineaments can be taken as an indirect evidence in favour to the theory of dilatncy. Nevertheless, it is necessary to make more detailed studies to make definitive conclusions.

\onecolumn

\begin{table}
\begin{center}
\begin{tabular}{|c|c|c|c|c|c|c|c|} \hline
Place & Date     & Magn.Mw.  & Depth & Lat.       & Long.     & Images available & Changes in lineaments \\ \hline 
Chile & 6/13/05 & 7.8            & 115.6  & -19.99   & -69.197 & -126, -69,  -5, +139 & yes \\
Chile & 9/17/03 & 5.8            & 127.1  & -21.467 & -68.325 & -138, -12, +86 & yes \\
Peru & 10/17/05 & 5.8           & 123     & -17.775 & -69.486 &-51, -3, +4, +132 & yes \\
Chile & 6/17/04 & 5.7 & 115.4 & -21.246 & -68.372 &-188, -35, -28, -19, +100   & yes \\
Peru & 01/27/04 & 5.2 & 120.1 & -17.69 & -70.65 & -128,-48,+73 & yes \\
China & 4/19/06 & 5.7 & 33.1 & 31.6 & 90.4 & -132, -116, -52, -20*, -4*  & ?  \\ \hline
\end{tabular}
\caption{Main characteristics of earthquakes analyzed}
\end{center}
\end{table}

\begin{figure}
\includegraphics[width=18cm]{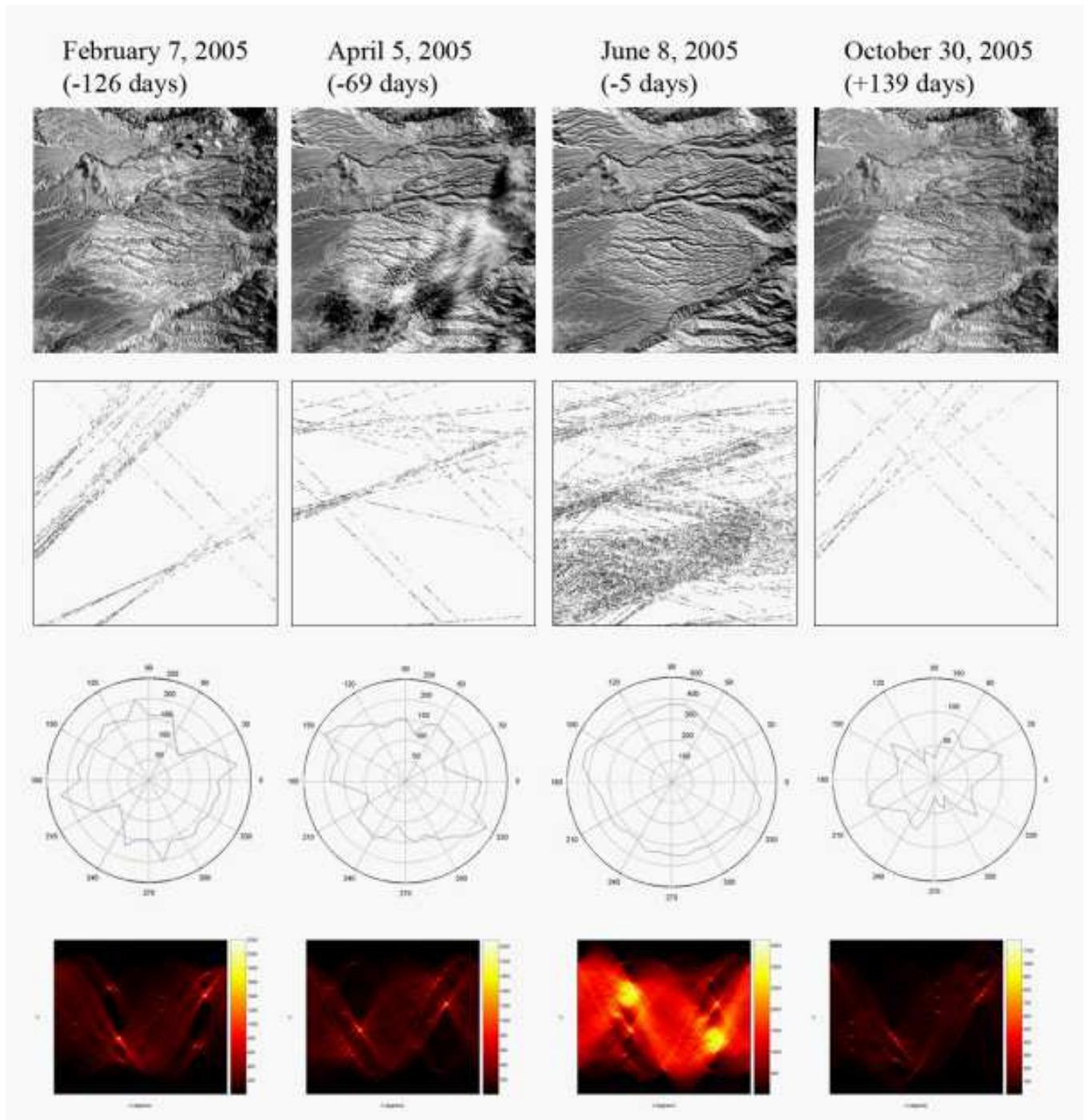} 

\caption{ From top to bottom: ASTER band 3 images around the June 13, 2005 earthquake (Chile, Arica). Systems of lineaments extracted from these images, using LESSA sofrware.}
\end{figure} 
    
\twocolumn
\section{Discussion and conclusions}

In this study we used the multispectral satellite images from ASTER/TERRA satellite for detection and analysis of lineaments in the areas around strong earthquakes with magnitude more than 5 Mw. A lineament is a straight or a somewhat curved feature in an image, which can be detected by a special processing of images, based on directional filtering and/or Hough transform. It was established that the systems of lineaments are very dynamical. By analyzing 5 events of strong earthquakes, it was found that a significant number of lineaments appeared approximately one month before an earthquake, and one month after the earthquake the lineament configuration returned to its initial state. These features were not observed in the test areas, situated hundreds kilometers away from the earthquake epicenters. 

The main question is how the lineaments extracted from images of 15-30 m (ASTER) in resolution are able to reflect the accumulation of stress deep in the crust given that the ground deformations associated with these phenomena are about a few centimeters?  The nature of lineaments is related to the presence of faults and dislocations in the crust, situated at different depth. If a dislocation is situated close to the surface, the fault appears as a clear singular lineament. In the case of a deep located fault, we observe the presence of extended jointing zones, easily detectable in satellite images even up to 200 m resolution. Nevertheless, how well lineaments can be detected strongly depends on a number of factors. In particular, it depends on the current level of stress in the crust. Generally, an enlarged presence of lineaments indicates that in these regions the crust is more permeable, allowing the elevation of fluids and gases to the surface. Accumulation of stress deep in the crust modifies all afore mentioned processes and leads to the variation in the density and orientationn of lineaments, previous to a strong earthquake.

\textbf{Acknowledgments}

We acknowledge Hiroji Tsu (Geological Survey of Japan CSJ)  ASTER Team Leader, Anne Kahle (Jet Propulsion Laboratory  JPL)  US ASTER Team Leader and the Land Processes Distributed Active Archive Center for providing the ASTER level 2 images. We acknowledge A. Zlatopolsky for providing the Lineament Extraction and Stripes Statistic Analysis (LESSA) software package and helpful suggestions. We thank very much Milton Rojas Gamarra for his assistance in image procesing. This work has been supported by DICYT/USACH grant.
\\
\\
\textbf{REFERENCES}

Abrams, M., The Advanced spaceborne Thermal Emission and Reflection Radiometer (ASTER): Data products for the high spatial resolution imager on NASA's Terra platform, International Journal of Remote Sensing, 21(5), 847-859, 2000.

  Aggarwal Y. P., Sykes L. R., Simpson D. W., Richards P. G., Space and temporal variations of ts/tp and P waves residuals at Blue Mountain Lake, J. Geophys. Res., 80, 718-732, 1973.

  Alparone, S., B. Behncke, S. Giammanco, M. Neri, E. Privitera, E., Paroxysmal summit activity at Mt. Etna (Italy) monitored through continuous soil radon measurements, Geoph. Res. Lett., 32(16), 10.1029/2005GL023352, 2005.

  Arellano-Baeza, A.; Zverev, A.; Malinnikov, V. Study of the structure changes caused by earthquakes in Chile applying the lineament analysis to the Aster (Terra) satellite data. 35th COSPAR Scientific Assembly, Paris, France, July 18-25, 2004. 

  Arellano-Baeza, A.; Zverev, A.; Malinnikov, Changes in geological faults associated with earthquakes detected by the lineament analysis of the ASTER (TERRA) satellite data., XI Latin American Symposium on Remote Sensing and Spatial Information Systems, Santiago, November 22-26, 2004.

  Arellano-Baeza A.A., A. Zverev, V. Malinnikov, Study of changes in the lineament structure, caused by earthquakes in South America by applying the lineament analysis to the Aster (Terra) satellite data, Advances in Space Research, doi:10.1016/j.asr.2005.07.068,  electronic access from 2005.

  Bak, P., C. Tang, and K. Wiesenfeld, Self-organized critucality, Physical Review A, 38(1), 364-374, 1988.

  Bolt, B. A., Earthquakes and Geological Discovery,  New York, Scientific American Library, 1993.

  Bondur, V.G., and A.T. Zverev, A method of earthquake forecast based on the lineament analysis of satellite images, Doklady Earth Sciences, 402, 561-567, 2005.

  Brune, J. N., Implications of earthquake triggering and rupture propagation for earthquake prediction based on premonitory phenomena. J. Geophys. Res., 84, 2195-2198, 1979.

  Campos, J., J. Dechabalier, A. Perez, P. Bernard, S. Bonvalot, M. Bouin, O. Charade, A. Cisternas, E. Clevede, V. Clouard, R. Dannoot, G. Gabalda, E. Kausel, D. Legrand, A. Lemoine, A. Nercessian, G. Patau, J. Ruegg, J. Vilotte, Source parameters and GPS deformation of the Mw 7.8 Tarapaca intermediate depth earthquake, abstract Nº S13B-0209, American Geophysical Union Fall Meeting, San Francisco, USA, 2005.

  Carver, G., J. Sauber, W. R. Lettis, R. C. Witter, Use of SRTM and Landsat-7 to evaluate seismic hazards, Kodiak Island, Alaska, abstract nnnnn 4513, EGS-AGU-EUG Joint Assembly, Nice, France, 6-11 April, 2003.

  Cervone, G., R.P. Singh, M. Kafatos, C. Yu, Wavelet maxima curves of surface latent heat flux anomalies associated with Indian earthquakes, Natural Hazards and Earth System Sciences, 5(1), 87-99, 2005.

  Comte, D., H. Tavera, C. David, D. Legrand, L. Dorbath, A. Gallego, J. Perez, B. Glass, H. Haessler, E. Correa, A. Cruz, Seismotectonic characteristics around the Arica bend, Central Andes (16S-20S): preliminary results, abstract nnnnnnnS41A-05, American Geophyisical Union Fall Meeting, San Francisco, USA, 2003.

  Cotilla Rodriquez, M. O., D. Cordoba Barba, Morphotectonics of the Iberian Peninsula, Pure and Applied Geophysics, 161(4), 755-815, 2004.

  Cross, A. M., Detection of circular geological features using the Hough Transform. International Jounal of Remote Sensing, 9, 1519-1528. 1988.

  David, C., D. Comte, H. Tavera, L. Audin, G. Herail, Crustal Seismicity and Recent Faults in Southern Peru, abstract nnnnnS43C-1014, American Geophysical Union, Fall Meeting, San Francisco, USA, 2004.		

  De Rubeis, V., P. Dimitriu, E. Papadimitriou, and P. Tosi, Recurrent patterns in the spatial behavior of Italian seismicity revealed by the fractal approach, Geoph. Res. Lett., 20, 1911-1914, 1993.

  Dey, S., S. Sarkar, R. P. Singh, Anomalous changes in column water vapor after Gujarat earthquake, Adv. Space. Res., 33(3), 274-278, doi: 10.1016/S0273-1177(03)00475-7, 2004.

  Fitton, N. C. and Cox S. J. D., Optimizing the application of the Hough Transform for the automatic feature extraction from geoscientific images. Computers and Geosciences, Vol. 24, P. 933-951, 1998.

  Funning, G. J., B. Parsons, T.J. Wright, J.A. Jackson, James E.J. Fielding, Surface displacements and source parameters of the 2003 Bam (Iran) earthquake from Envisat advanced synthetic aperture radar imagery, Surface displacements and source parameters of the 2003 Bam (Iran) earthquake from Envisat advanced synthetic aperture radar imagery, 110(B9), doi: 10.1029/2004JB003338, 2005. 

  Gokhberg, M. B., O. A. Pokhotelov, V. A. Morgounov, Earthquake prediction seismoelectromagnetic phenomena, London, Taylor and Francis Ltd, ISBN/CatNo: 2881249213, 208 p, 1995.

  A. A. Griffiths, The phenomenon of rupture and flow in solids, Phil. Trans. Roy. Soc. London A 221, 163-198, 1921.
  Griggs, D. T., D. D. Jackson, L. Knopoff, and R. L. Shreve, Earthquake prediction: Modeling the anomalous Vp/Vs source region, Science, 187, 537-540, 1975.

  Hobbs, W. H., Lineaments of the Atlantic border region, Geological Society American Bulletin, 15, 483-506, 1904.

  Kagan, Y.Y., and D.D. Jackson, New seismic gap hypothesis: Five years after, J. Geoph. Res., 100(B3), 3943-3959, 1995.

  Karnieli, A., A. Meisels, L. Fisher, Y. Arkin, Automatic Extraction and Evaluation of Geological Linear Features from Digital Remote Sensing Data Using a Hough Transform. Photogrammetric Engineering and Remote Sensing. 62(5), 525-531, 1996.

  Keilis-Borok V. I. The litosphere of the Earth as nonlinear system with implications for earthquake prediction, Rev. Geophys., 28(1), 5-34, 1990.

  King, G. C. P. The accommodation of large strains in the upper lithosphere of the earth and other solid by self-similar fault system: the geometrical origin of the b-value, PAGEOPH. 121, 567-585, 1983.

  Koike, K., Nagano S.  and Kawaba K., Construction and Analysis of Interpreted Fracture Planes through Combination of Satellite-Image Derived Lineaments and Digital Elevation Model Data. Computers and Geosciences, 24, 573-583, 1998.

  Koizumi, N., Y. Kitagawa, N. Matsumoto, M. Takahashi, T. Sato, O. Kamigaichi, K. Nakamura, Preseismic groundwater level changes induced by crustal deformations related to earthquake swarms off the east coast of Izu Peninsula, Japan, Geoph. Res. Lett., 31(10), doi: 10.1029/2004GL019557, 2004.

  Lasserre, C., G. Peltzer, F. Cramp� Y. Klinger, J. Van der Woerd, P. Tapponnier, Coseismic deformation of the 2001 Mw = 7.8 Kokoxili earthquake in Tibet, measured by synthetic aperture radar interferometry, J. Geoph. Res., 110(B12), doi: 10.1029/2004JB003500, 2005.

  Legrand, D., A. Cisternas, L. Dorbath, Multifractal an�isis of the 1992 Erzincan aftershock sequence, Geophys. Res. Lett., 23(9), 933-936, 1996.	

  Legrand, D., Co-seismic deformation of the crustal Mw=6.3, 2001, Chusmiza, Chile event triggered by the subduction Mw=8.4, 2001, Arequipa, Peru earthquake studied using seismological data, abstract S43A-1051, American Geophysical Union, Fall Meeting, San Francisco, USA,  2005.

  Mah, A., Taylor G. R., Lennox P., and Balia L., Lineament Analysis of Landsat TM images, Northern Territory, Australia. Photogrammetric Engineering and Remote Sensing, 61, 761-773, 1995.

  Mjachkin, V.I., W. Brace, G.A. Sobolev, and J. Dieterich, Two models for earthquake forerunners, Pure Appl. Geophys. 113, 169-180, 1975.

  Mjachkin, V.I., B.V. Kostrov., G.A. Sobolev, O.G. Shamina, Fundaments of the physics of the earthquake epicenter and the earthquake precursors, Moscow, Nauka, p.6-29, 1975. (In Russian).

  Morris, K., Using knowledge-base rules to map the threedimensional nature of geological features. Photogrammetric Engineering and Remote Sensing, 57, 1209-1216, 1991.

  Morrow, C., W.F. Brace, Electrical resistivity changes in tuffs due to stress, J. Geoph. Res., 86(B4), 2929-2934, 1986.
  O'Leary, D. W. Friedman, J. D. and Pohn, H. A., Lineament, linear lineation some proposed new standards for old terms. Geological Society America Bulletin, 87, 1463-1469., 1976. 

  Ostapenko V.F., V.A. Krasnoperov, Analysis of natural neutron flux in a seismically active zone, Natural Hazards and Earth System Sciences, 3, 777-780, 2003.

  Ouzounov, D., F. Freund, Mid-infrared emission prior to strong earthquakes analyzed by remote sensing data, Adv. Space Res., 33(3), 268-273, doi: 10.1016/S0273-1177(03)00486-1, 2004.

  Pulinets, S. A., A.D. Legen'ka, T.V. Gaivoronskaya, V.Kh. Depuev, Main phenomenological features of ionospheric precursors of strong earthquakes, J. Atm. Sol.-Terr. Phys., 65(16-18), 1337-1347, doi:10.1016/j.jastp.2003.07.011, 2003.

  Reid, H. F., The mechanics of the earthquake, v. 2 of The California earthquake of April 18, 1906: Report of the State Earthquake Investigation Commission: Carnegie Institution of Washington Publication 87, 1910.	

  Sammis, C. G., and D. Sornette, Positive feedback, memory, and predictability of earthquakes, Proceedings of the National Academy of Sciences, 99, 2501-2508, 2002.

  Satyabala, S. P., Coseismic ground deformation due to an intraplate earthquake using synthetic aperture radar interferometry: The Mw6.1 Killari, India, earthquake of 29 September 1993, J. Geoph. Res., 111(B2), doi: 10.1029/2004JB003434, 2006.	

  Sharma, D. K., M. Israil, R. Chand, J. Rai, P. Subrahmanyam, S. C. Garg, Signature of seismic activities in the F2 region ionospheric electron temperature, J. Atm. Sol.-Terr. Phys., 68(6), 691-696, doi: 10.1016/j.jastp.2006.01.005, 2006.

  Scholz, C. H., Experimental study of structuring process in brittle rocks, J. Geophys. Res., 73(4), 1447-1454, 1968.

  Scholz, C. H., L. R. Sykes, and Y. P. Aggarwal, Earthquake prediction: A physical basis, Science, 181, 803-809, 1973.

  Serebryakova, O. N., S. V. Bilichenko, V. M. Chmyrev, M. Parrot, J. L. Rauch, F. Lefeuvre, O. A. Pokhotelov, Electromagnetic elf radiation from earthquake regions as observed by low-altitude satellites, Geophys. Res. Lett., 19(2), 91-94, 1992. 

  Singh, R. P., and D. Ouzounov, Earth processes in wake of Gujarat earthquake reviewed from space, EOS Transactions, AGU, 84(26), 244-244, doi: 10.1029/2003EO260007, 2003.

  Singh, V.P., and R.P. Singh, Changes in stress pattern around epicentral region of Bhuj earthquake, Geoph. Res. Lett., 32, doi: 10.1029/2005GL023912, 2005.

  Schmidt, D. A., R. Brgmann, InSAR constraints on the source parameters of the 2001 Bhuj earthquake, Geoph. Res. Lett., 33(2), doi: 10.1029/2005GL025109, 2006.

  Smirnov V. B., A. V. Ponomarev, A. D. Zavialov, Structure of the acoustical regimen of seismic processes in the samples of rocks, Physics of the Earth (Fisika Zemly), 31(1), 38-58, 1995. (In Russian).

  Sobolev, G.A. A.A. Semerchan, B.G. Salov, etc., Precursor of the destruction of big blocks of rock, Izvestiya, Physics of the Solid Earth, No 8, 29-43, 1982. 

  Sobolev G. A., Babichev O. V., Los V. F., et. al., Precursors of the destruction of water-containing blocks of rocks, J. Earthquake Prediction Res., (1), 63-91, 1996.

  Sobolev, G. A., and A. V. Ponomarev, Physics of earthquakes and its precursors, M. Nauka, ISBN 5-02-002832-0, 270 p., 2003 (in Russian).

  Stich, D., G. Alguacil, J. Morales, The relative locations of multiplets in the vicinity of the Western Almeria (southern Spain) earthquake series of 1993-1994, Geophysical Journal International, 146(3), 801-812, 2001. 

  Szen, M. L. and Toprak V., Filtering of satellite images in geological lineament analyses: an application to a fault zone in Central Turkey. International Journal of Remote Sensing, Vol. 19, P. 1101-1114, 1998.

  Trigunait, A., M. Parrot, S. Pulinets, F. Li, Variations of the ionospheric electron density during the Bhuj seismic event, Ann. Geoph., 22(12), 4123-4131, 2004. 

  Varnes, D. J., Predicting earthquakes by analyzing accelerating precursory seismic activity, PAGEOPH. 130(4), 661-686, 1989.
  Wang, J., Howarth, P.J.: Use of the Hough Transform in Automated Lineament Detection. IEEE Tran. Geoscience and Remote Sensing, Vol. 28. No. 4. P. 561-566, 1990.

  Whitcomb, J. H., J. D. Garmany, and D. L. Anderson, Earthquake prediction: variation of seismic velocities before the San Fernando earthquake,  Science, 180, 632-641, 1973.

  Zlatopolsky, A. A., Program LESSA (Lineament Extraction and Stripe Statistical Analysis): automated linear image features analysis  experimental results, Computers and Geosciences, 18(9), 1121-1126, 1992.

  Zlatopolsky, A. A., Description of texture orientation in remote sensing data using computer program LESSA, Computers and Geosciences, 23(1), 45-62, 1997.

\end{document}